# The Effects Of Longitudinal And Circumferential Cracks On The Torsional Dynamic Response Of Shafts

Mohsen Nabian, Hamid.N Hashemi

*Abstract*

Turbo generators shafts are manufactured through the extrusion process. This results in formation of weak planes along the extrusion process. It has been observed that large longitudinal cracks often form in these shafts before any circumferential cracks when these shafts are subjected to cyclic torsion due to electrical line faults. The presence of these cracks could severely compromise the shaft resonance frequencies. Dynamic response of shafts with longitudinal and circumferential cracks is investigated. The longitudinal cracked section of the shaft section is modeled as a shaft with reduced torsional rigidity. The torsional rigidity is obtained as a function of the crack depth. It was found for various shaft diameters, torsional rigidity could be represented as a function of crack depth/ shaft radius only. The circumferential cracked section is modeled as a torsional spring. The torsional spring constant has been obtained using fracture mechanics. It was found the resonance frequency of the shaft may be little affected by the presence of longitudinal crack. The resonance frequencies of the shaft with the circumferential crack depend on the crack length and location. The effects of crack surface interactions for both longitudinal and circumferential cracks were also investigated.

## INTRODUCTION

Turbo-generator shafts are manufactured using high strength alloy steels through the extrusion process. Microstructures of these alloys contain carbides and other inter-metallic compounds which line up along the shaft resulting in an anisotropic microstructure with weak plane along the shaft axis. During normal operation these shaft are subjected to sub – synchronous resonance. However electrical faults results in shaft to be subjected to high cyclic torsion often causing its plastic deformation. It has been observed that during such incidents, large longitudinal cracks forms in the direction of shaft before circumferential cracks develop. These longitudinal cracks often propagate at higher crack growth rates than the circumferential crack due to having propagation in the weaker plane of the material [1-5]. Figure 1 shows formation of both longitudinal and circumferential crack in a specimen subjected to cyclic torsion. Furthermore in contrast to the circumferential crack growth which results in a complex fracture surface resembling factory roof, longitudinal cracks form a smooth fracture surface. Crack interaction in the circumferential crack growth reduces the effective stress intensity factor at the crack tip. This reduction in the stress intensity factor is almost absent in the longitudinal crack, except for small frictional interaction between crack surfaces. Both longitudinal and circumferential cracks change the resonance frequencies of the shaft and may further reduce its life expectancy. Vaziri and Hashemi [6-7] in a series of investigations studied dynamic response of shafts with circumferential crack considering complexity of the cracked surface. The effect of crack surface morphology (factory roof, pitch height and wave length) on the effective stress intensity factor and energy loss was investigated. It was shown that effective stress intensity factor depends on the applied stress intensity factor and fracture surface morphologies. The crack surface interactions diminish at high applied stress intensity factor. The loss factor associated with interaction of cracked surfaces and crack tip cyclic plasticity was also investigated. It was shown, that for small applied stress intensity factor, crack interaction dominates the energy loss and for high applied stress intensity factor, plasticity dominates the energy loss. The results showed that the natural frequency of the circumferentially cracked shaft decreases with the length of the crack, however if the cracked surface interaction is sever, the natural frequency approaches to that of uncracked shafts [7]. Despite of this body of information, there have been few works investigating the effect of the longitudinal cracks on the dynamic behavior of shafts and various other structures. Maselko has analytically and numerically studied the dependence of the opening of the longitudinal crack on the geometric and physio-mechanical parameters of a shell subjected to an internal pressure [8]. Several literatures have concentrated on the modeling of circumferential cracks. A crack has been modeled by a spring [9, 10], elastic hinge [11,12], cut-out [13], a pair of concentrated couples [14,15], a zone with a reduced Young's modulus [16,17], or its effect has been taken into account by semi-empirical functions describing stress and strain distribution by the volume of the cracked material [18]. The changes in the natural frequencies [19, 20], mode shapes of vibrations [21-23,30], the measured dynamic flexibility [24,25] and the amplitudes of forced response [26-29] could provide information regarding heath status of the turbo-generator shafts with longitudinal crack.

The purpose of this investigation is to understand the effect of longitudinal crack on the natural frequency of the shaft and to understand how the natural frequencies changes when both longitudinal and circumferential crack are present in a shaft [31].

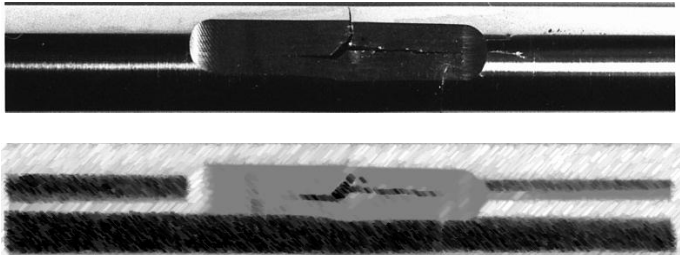

Fig. 1. formation of circumferential and longitudinal crack from a defect in shaft subjected to a torsional loading

1. **Theoretical and Numerical Investigations**

*2.1 Longitudinal Crack*

*2.1.1 Torsional rigidity of a round shaft with a longitudinal crack.*

The closed form solution for the torsional rigidity of a shaft with circular, elliptical, and triangular cross section is obtained using Prandtl stress functions [31]. For other shaft geometries, numerical solution such as finite element analysis is required to obtain its torsional rigidity. For a shaft with a longitudinal crack, ABAQUAS finite element program was used to find its torsional rigidity. Figure 2 shows the shaft geometry and its boundary condition. The shaft was fixed at one end and was attached to a rigid plate at the other end. This prevented its cross section to warp. In a shaft with a longitudinal crack section, the uncracked section prevents warping of the cracked region. For this reason, the finite element model considered rigid plate at one end. The model was subjected to torsion to the other end while the rotation of rigid plate was monitored.

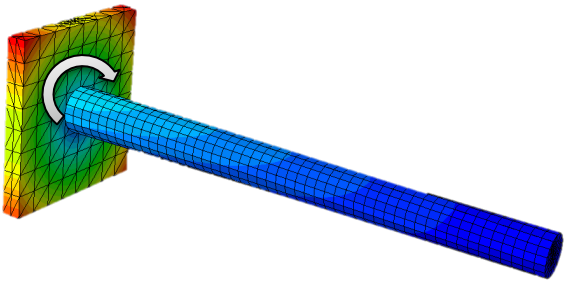

Fig. 2. Finite element model of a shaft with a longitudinal crack subjected to torsion

The torsional rigidity of the cracked shaft was obtained from

$$\quad (1)$$

Where the $T$ is the applied torque, $GJ^*$ is the torsional rigidity, $\frac{\partial \theta}{\partial x}$ is the rotation gradient along the shaft. The torsional rigidity of the cracked shaft was related to the torsional rigidity of uncracked shaft by defining

$$GJ^* = GJ \times f \quad (2)$$

The result shows that for various shaft diameter the correction factor $f$ is only function of $\frac{C_L}{R}$, Figure 3.

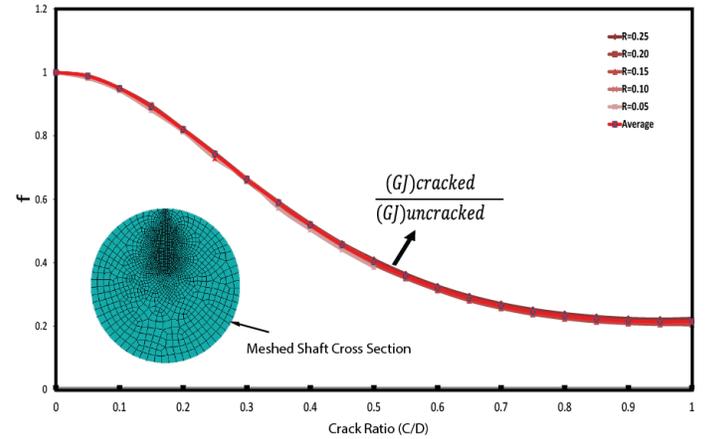

Fig. 3 Torsional rigidity correction factor for a shaft with longitudinal crack

The correction factor was related to the crack depth and the shaft radius as,

$$f = f\left(\frac{C_L}{R}\right) = 3.76\,(C_L/R)^5 - 12.82\left(\frac{C_L}{R}\right)^4 + 16.16\left(\frac{C_L}{R}\right)^3 \\ - 8.05\left(\frac{C_L}{R}\right)^2 + 0.15\left(\frac{C_L}{R}\right) + 1 \quad (3)$$

For torsional vibration analysis of shafts with longitudinal crack, the shaft was modeled as an equivalent shaft with the radius R and torsional rigidity of $GJ \times f\left(\frac{C_L}{R}\right)$, Fig. 4.

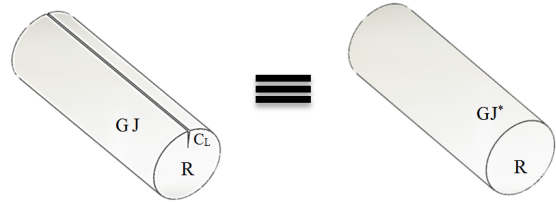

Fig. 4. Equivalent model for a shaft with a longitudinal crack subjected to torsion

*2.1.2 Loss factor modeling of Longitudinal cracks:*

The effect of energy loss due to the crack surface interactions, can be incorporated by considering $G$ as a complex number, $G(1 + i\eta_L)$, where $\eta_L$ is the loss factor associated with the crack surface interaction. The shear displacement along the crack at each cross section is presented by Gross and Mendelsohn (1988) as

$$u_3 = \frac{2k_{III}}{G\sqrt{\pi C}} (C^2 - x^2)^{0.5} \qquad (4)$$

where $G$ is the shear modulus of the shaft material, $C$ is the crack length and $k_{III}$ is the applied mode III stress intensity factor which is obtained from (Teda et al.,2000).

$$k_{III} = \frac{G\frac{d\varphi}{dz}C\,(2R-C)^2}{4\pi(R-C)^2} \left\{ \frac{4R(R-C)+3C^2}{2C\sqrt{R(R-C)}} \arctan\left(\frac{2\sqrt{R(R-C)}}{C}\right) - 3 \right\}$$
$$* \sqrt{\frac{2\pi CR}{2R-C}} \qquad (5)$$

where $\frac{d\varphi}{dz}$ is the rotation gradient along the shaft.

For simplicity, $k_{III}$ can also be rewritten as
$$k_{III} = G\, c_1(C,R)\, \frac{d\varphi}{dz} \qquad (6)$$

by substituting equation 6 into 4,
$$u_3 = \frac{2\, C_1(a,R)\frac{d\varphi}{dz}}{\sqrt{\pi a}} (a^2 - x^2)^{0.5} \qquad (7)$$

The interaction between the mutual crack surfaces creates shear stress along the surfaces. By assuming a constant shear stress profile as,

$$\tau = \alpha\, \tau_y \qquad (8)$$

the energy loss per cycle in the model due to the crack surface interaction can be obtained from,

$$\Delta W = 4 \int_0^a (\alpha\, \tau_y\, dx)(u_3) \qquad (9)$$

Thus, the energy loss is derived as,

$$\Delta W = 2\sqrt{\pi}\; a^{1.5}\, \alpha\, \tau_y\, C_1(a,R) \left(\frac{d\varphi}{dz}\right)_{max} \qquad (10)$$

In addition, the total potential energy due to the torsion of the shaft per cycle is obtained as,
$$W = \int \frac{1}{2G}(\tau^2)\, d\vartheta = \frac{1}{2G}\left(G\,r\,\frac{d\varphi}{dz}\right)^2 dA dx = \frac{G}{2} J^*\left(\frac{d\varphi}{dz}\right)^2 \qquad (11)$$

By the definition of energy loss factor as the ratio of energy loss per total potential energy of the structure in each cycle,

$$\eta = \frac{\Delta W}{2\pi\, W_{max}} \qquad (12)$$

The energy loss factor $\eta$ can be formulated in the following form,

$$\eta = \frac{2\, a^{1.5}\, \alpha\, \tau_y\, C_1(a,R)}{\sqrt{\pi}\, G\, J^* \left(\frac{d\varphi}{dz}\right)_{max}} \qquad (13)$$

*1.2 Shaft with Circumferential crack*

In order to evaluate the torsional dynamic response of the shaft with circumferential crack, The shaft is modeled as having two segments connected by a torsional spring and a torsional damping, Fig. 5 The torsional stiffness and damping constants of the spring are evaluated by considering the energy loss at the crack region due to the cyclic plasticity deformation at the crack tip and the frictional energy loss due to the crack surface interaction using fracture mechanics. The detail of this analysis is presented in [6-7].

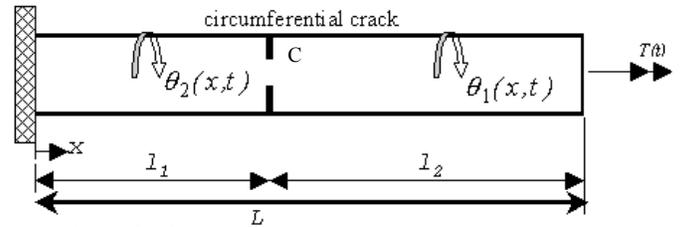

Fig.5 Schematic model of a shaft with a circumferential crack.

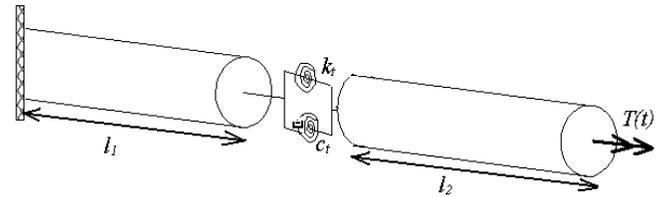

Fig. 6. The physical model corresponding to Fig. 5

The torsional spring is defined as,

$$K_c = K_t(1 + i\eta_c) \qquad (14)$$

where $K_t$ is the torsional stiffness and $\eta_c$ is the circumferential crack loss factor.

The torsional stiffness $K_t$ is computed from:

$$K_t = GR^3 M(\gamma) \qquad (15)$$

where

$$M(\gamma) = \frac{\pi \gamma^{2.5}}{4(1-\gamma)[1 + \frac{1}{2}\gamma + \frac{3}{8}\gamma^2 + \frac{5}{16}\gamma^3 + \frac{32}{128}\gamma^4 + 0.208\gamma^5]} \qquad (16)$$

and

$$\gamma = \frac{a}{R} \qquad (17)$$

where $a$ is the radius of uncracked section.

The loss factor of circumferentially cracked shaft was obtained by considering crack surface interactions [6-7]. Here we have used a range of loss factor to represent the severity of crack surface interactions. For the longitudinal crack, we have also considered loss factor due to crack surface interactions. However, longitudinal crack surface interactions are relatively less pronounced compared to the circumferential crack. Here we have also considered a range of loss factor for the longitudinal crack to understand its effects on dynamic response of the shaft.

*2.3 Dynamic response of a shaft with both circumferential and longitudinal cracks*

The torsional dynamic response of the cracked turbo-generator shaft subjected to a harmonic torsional loading at its free end, $T_0 e^{i\omega t}$, is evaluated considering the equation of motion for each segment in Fig (7) and their corresponding boundary conditions.

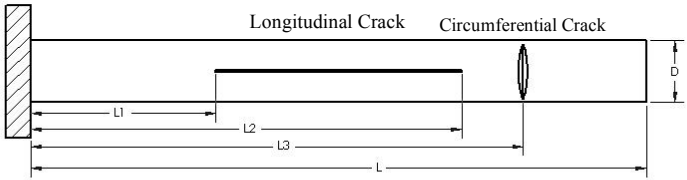

Fig. 7. Schematic model of a shaft with both longitudinal and circumferential cracks.

The equation of motion for the uncracked section can be presented as

$$G \frac{\partial^2 \theta}{\partial x^2} = \rho \frac{\partial^2 \theta}{\partial t^2} \qquad (18)$$

Where $G$ and $\rho$ are the material shear modulus and density respectively, and $\theta$ is the shaft cross section rotation. The equation of motion for the shaft with a longitudinal crack is presented as

$$G(1 + i\eta_L) f\left(\frac{C_L}{R}\right) \frac{\partial^2 \theta}{\partial x^2} = \rho \frac{\partial^2 \theta}{\partial t^2} \qquad (19)$$

The steady state displacement function $\theta_i(x,t)$ corresponding to each model segment when the shaft is subjected to harmonic torsional loading is of the type,

$$\theta_i(x,t) = \bar{\theta}_i(x) e^{i\omega t} \qquad (20)$$

The governing equation to find the displacement $\bar{\theta}_i(x)$ can be found by substituting Eq. 10 into the Equ. 8. The resulting Equation can be simplified by introducing non-dimensionalized parameters,

$$\varepsilon = \frac{x}{L} \qquad (21)$$

and

$$a_i = \frac{\rho L^2 \omega^2}{G_i} \qquad (22)$$

as

$$\frac{d^2 \bar{\theta}_i}{d\varepsilon^2} + a_i \bar{\theta}_i = 0 \qquad (23)$$

Solution of Eq. (13) can be written in the following form:

$$\bar{\theta}_i(\varepsilon) = \sum_{j=1}^{2} A_{ij} e^{S_{ij}\varepsilon} \qquad (24)$$

Where $S_{i1} = i\sqrt{a_i}$ and $S_{i2} = -i\sqrt{a_i}$ and $A_{ij}$ are constants which are found from the boundary conditions of each segment. The boundary conditions corresponding to the shaft shown in the Fig. 7 are presented as,

$$\begin{aligned}
&\text{at } \varepsilon_1 = 0 & &\bar{\theta}_1(\varepsilon_1) = 0 \\
&\text{at } \varepsilon_2 = L_1/L & &\bar{\theta}_1(\varepsilon_2) = \bar{\theta}_2(\varepsilon_2) \\
&\text{at } \varepsilon_2 = L_1/L & &T(\varepsilon_2^-, t) = T(\varepsilon_2^+, t) \\
&\text{at } \varepsilon_3 = L_2/L & &\bar{\theta}_1(\varepsilon_3) = \bar{\theta}_2(\varepsilon_3) \\
&\text{at } \varepsilon_3 = L_2/L & &T(\varepsilon_3^-, t) = T(\varepsilon_3^+, t) \\
&\text{at } \varepsilon_4 = L_3/L & &T(\varepsilon_4^+, t) = T(\varepsilon_4^-, t) \\
&\text{at } \varepsilon_4 = L_3/L & &T(\varepsilon_4^+, t) = K_t(\theta(\varepsilon_4^+, t) - \theta(\varepsilon_4^-, t)) \\
&\text{at } \varepsilon_5 = 1 & &T(\varepsilon_5, t) = T_0 e^{i\omega t}
\end{aligned} \qquad (25)$$

where $T(\varepsilon, t)$ can be related to $\theta(\varepsilon, t)$ using Eq.(1).

2. **Results and Discussions**

The results are provided for 4140 steel (in annealed condition) shaft of 2m long and 0.1m radius. Table 1 shows the mechanical properties of 4140 steel.

| Material | Elastic Modulus |
|---|---|
| 4140 steel, in annealed condition | 200Gpa |
| Density | Yield Stress |
| 7850 kg/m³ | 420Mpa |
| Poisson's ratio | |
| 0.30 | |

Table 1. Material properties of models

In order to understand the effect of each crack on the dynamic response of the shaft, a comprehensive study has been conducted and presented for (1) a shaft with a longitudinal crack (2) for a shaft with a circumferential crack, and (3) when both cracks are present.

*Case 1) longitudinal crack*

### a) The effect of length and depth of the crack

Assuming the longitudinal crack is located in the middle of the shaft and crack surface interaction is minimal ($\eta_L = 0$) a MATLAB-based codes were developed to analyze the dynamic response of the cracked shaft. Figs. 8 and 9 show the effect of crack length and depth on the first resonant frequency. The results indicate that both crack length and crack depth could have a significant effect on the natural frequency of the shaft. For the small crack depth ($c/R = 0.1$), the longitudinal crack length has apparently negligible effect on its natural frequency. This is even true if the entire shaft is cracked longitudinally. The effect of crack depth is more pronounced when $c/R$ is greater than 0.3.

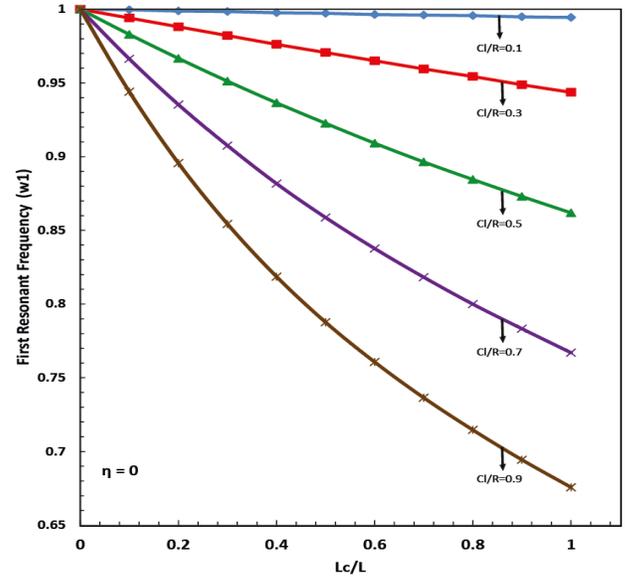

Fig.9. The effect of crack depth on the 1st resonant frequency of the shaft for various crack depth ratio.

### b) The effect of surface interaction

The effect of surface interaction is taken into account in the modeling equations by defining the term energy loss factor. According to the corresponding equations for loss factor, this value is influenced by the applied torque and is expected to be decreased as the applied torque is increased. The variation of crack surface interaction over a wide range of longitudinal crack depth is indicated in Figure 10 for a specific applied torque T. The variation of energy loss factor strongly depends on the shear stress distribution approximation. However, even with the strictest shear stress distribution approximation ($\alpha = 0.9$), energy loss factor can only increase up to 1.2 for crack depth ratio (C/R) 1, while, the corresponding loss factor value for circumferential cracks can rises up to the order of 20 to 50 for the same crack depth size. However, small loss factor values for the longitudinal crack region can have significant effect on the dynamic response of the shaft since the longitudinal crack occures along a considerable length of the shaft.

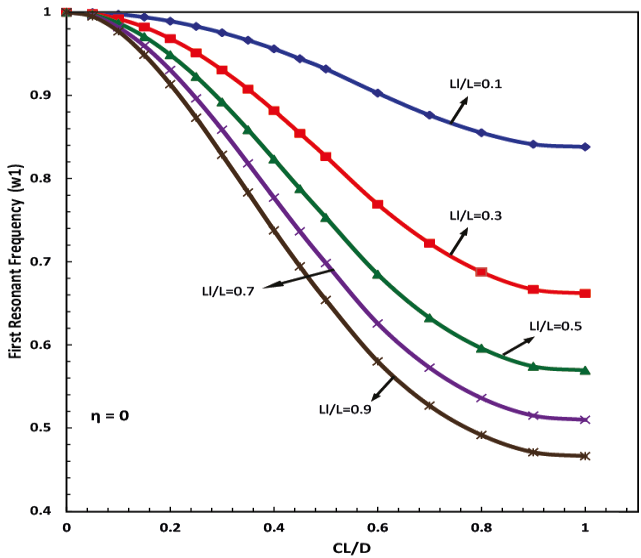

Fig.8. The effect of crack depth on the 1st resonant frequency for various crack length ratio.

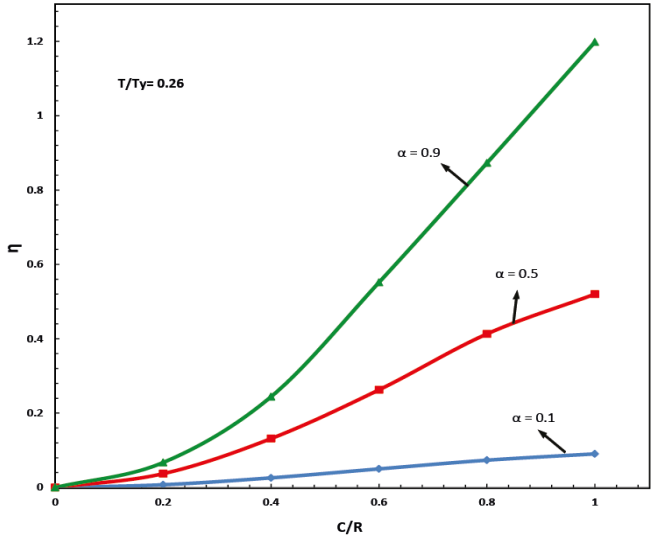

Fig.10. The variation of crack loss factor over a wide range of crack depth for different stress distributions, $\alpha = 0.1$, 0.5 and 0.9. The load $T/T_y$=0.26 and $T_Y$ is the required torque to generate yield stress in the torsional shaft.

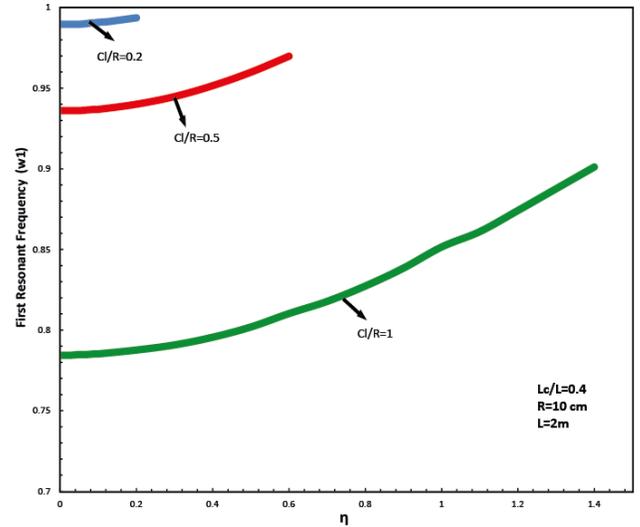

Fig.12. Effects of the energy loss of the longitudinal crack on the frequency response of the shaft for different crack depth ratios. The applied load $T/T_y$ =0.26 and $T_Y$ the require torque to generate yield stress in the torsional shaft.

The effect of energy loss factor on the peak response frequency of the structure is represented in Figures 14 and 15.. The results clearly show the variation of frequency changes for each specific crack depth. For the small cracks, the surface interactions have little effect on the dynamic response of the structure, while for higher crack depth ratios, peak resonance frequency may be affected by 5% to 15%.

*c) Mode shapes*

Assuming the longitudinal crack is located in the middle of the shaft with the length Lc/L=0.3, the effect of crack depth on the first and second mode shape of the shaft are shown in Figure 13 and 14 . Although the mode shapes of the structure has been significantly affected by the crack properties, the general trend of the mode shape curves are still unchanged.

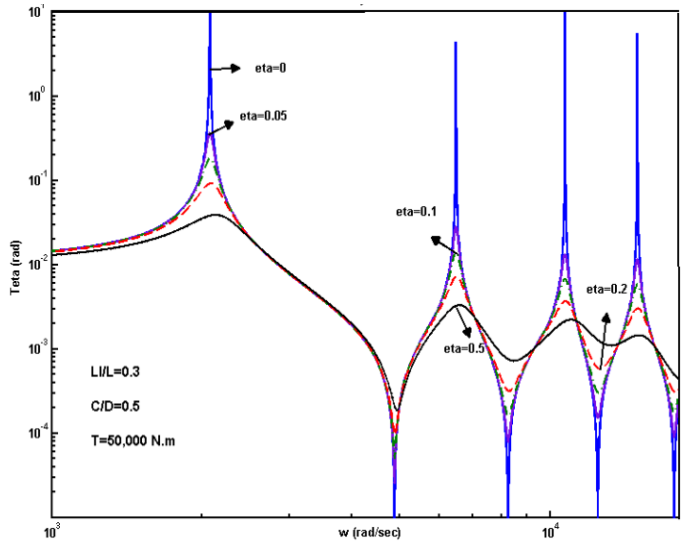

Fig.11. Effects of energy loss on the frequency response of the shaft with longitudinal crack depth C/D=0.5, length Ll/L=0.3 and applied load 50,000 N.m.

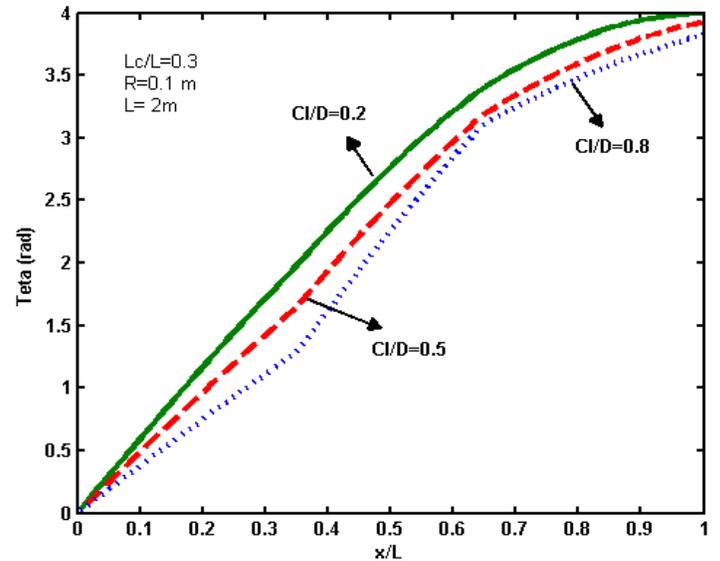

Fig.13. effects of longitudinal crack depth on the 1$^{st}$ mode shape of the shaft with crack length Lc/L=0.3.

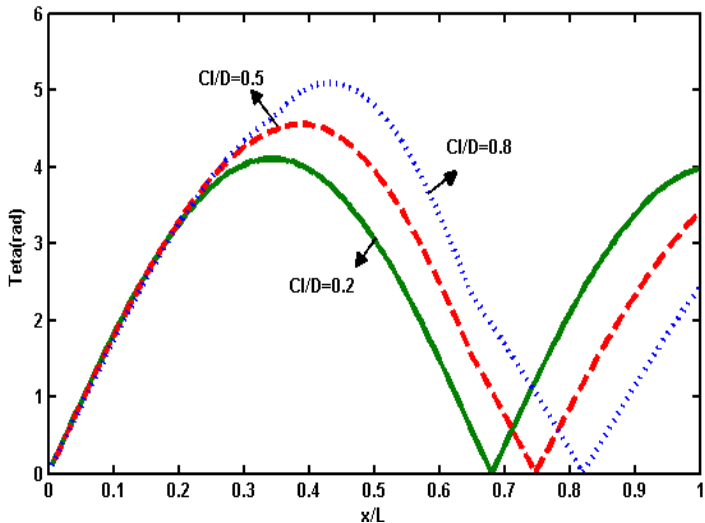
Fig.14. effects of longitudinal crack depth on the 2$^{nd}$ mode shape of the shaft with crack length Lc/L=0.3.

### d) Crack location detection

Although, the location of the longitudinal crack is not identified or hardly identified by the forced vibration response of the shaft, the results indicate that the derivates of the forced vibration response of the shaft in respect to x can determine the possible location of the longitudinal crack in the shaft, Figure 15. The second derivatives clearly shows the location of the crack as there exhibits a sharp change in the curve. The higher derivatives of the response magnify the location of the crack.

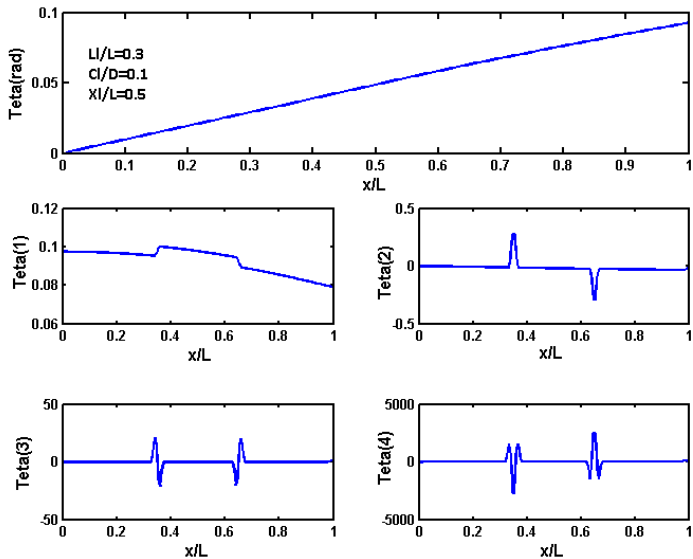
Fig.15. Forced vibration response of a longitudinally cracked shaft and its first four derivate with respect to x. The crack is located in the middle of the shaft and its length ratio and depth ratio are Ll/L=0.3 and Cl/D=0.1.

### e) Effect of concentrated mass

The effect of the concentrated mass and its location on the natural frequencies of the longitudinally cracked shaft is examined. Assuming the crack occurred in the middle of the shaft with the crack length Ll/L=0.3, the effect of concentrated mass located in the left side (clamped end side) of the crack and right side (free end side) of the crack on the first resonant frequency of the shaft is graphically represented in figure 16 and 17 respectively. From these plots, it is observed that for lower mass ratios with mass located in the clamped end side of the shaft, crack properties can strongly influence the dynamic response of the shaft, however, the higher concentrated mass ratios (m/M > 1.2 ) can vanish the effect of longitudinal crack, figure 16. But, in the shaft case with concentrated mass located in the free end side, the magnitude of mass ratio has little or no effect on the dependency of the vibration response of the shaft from the longitudinal crack properties.

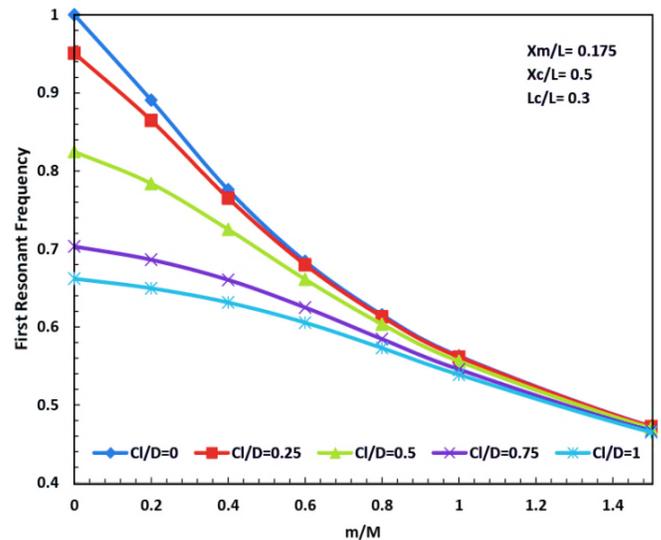
Fig.16. Effects of concentrated mass on the first resonant frequency of a shaft with longitudinal crack, the mass is located at Xm/L=0.175. M is denoted as the mass of the shaft.

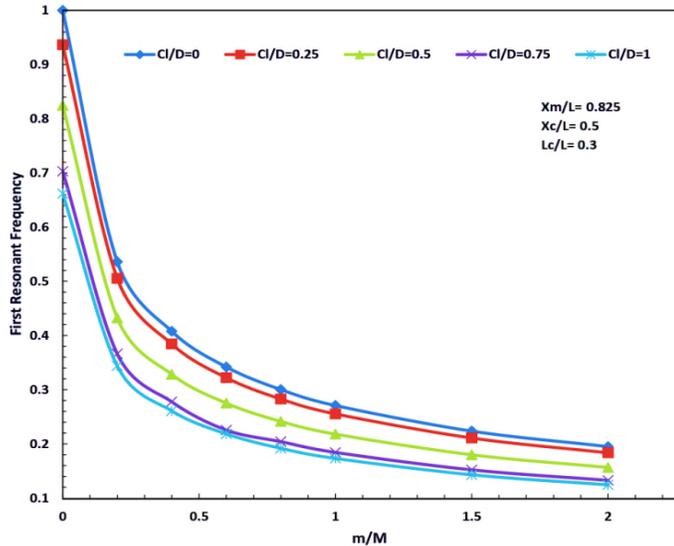

Fig.17. Effects of concentrated mass on the frst resonant frequency of a shaft with longitudinal crack, the mass is located at Xm/L=0.825.

As expected, the location of the concentrated mass can strongly affect the dynamic behaviour of the shaft. A comprehensive information about the effect of mass location on the first resonant frequency of the structure for various crack depth size is depicted in figure 17. The results indicate that the vibration response of the shaft with the concentrated mass located near the clamped end is strongly influenced by the crack size, while crack parameters becomes less dominant as the concentrated mass approaches the crack .On the other side of the crack (free end side) the resonant frequency drops uniformly as the mass is moved to the free end. This trend is aproximately the same for the various crack depth sizes.

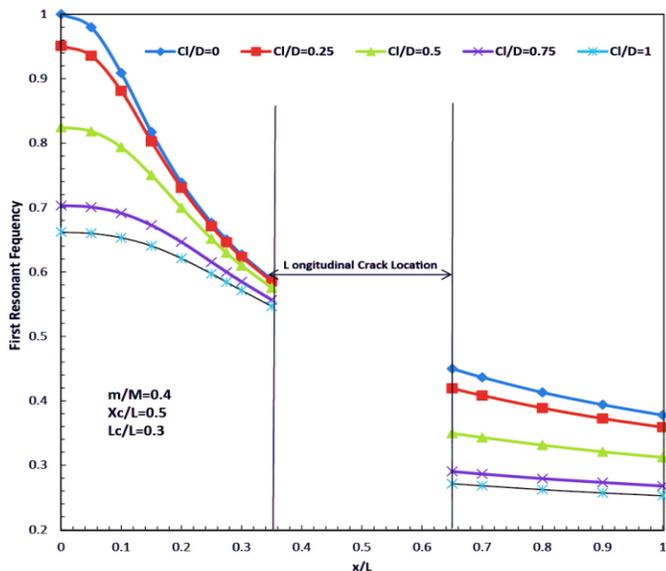

Fig.17. Effects of concentrated mass location on the frst resonant frequency of a shaft with longitudinal crack. The concentrated mass value is m/M=0.4, the crack is located in the middle of the shaft and its length is Lc/L=0.3.

## Case 2) circumferential crack

Assuming there is only one circumferential crack in the shaft, the results shown in Figs, 11 and 12 indicate that both crack location and its depth have a significant effect on the shaft peak frequency response. For a crack located near its free end $(x/L > .7)$ the frequency response of the shaft is not altered compared to uncracked shaft when $c/R < 0.6$. The effect of the crack surface interactions is also shown in Fig. 12. Here for a sever crack surface interactions, the peak response frequencies approaches to the natural frequencies of uncracked shaft. The peak frequency response for the circumferential cracked shaft is lower than the one with a longitudinal crack, Fig. 12.

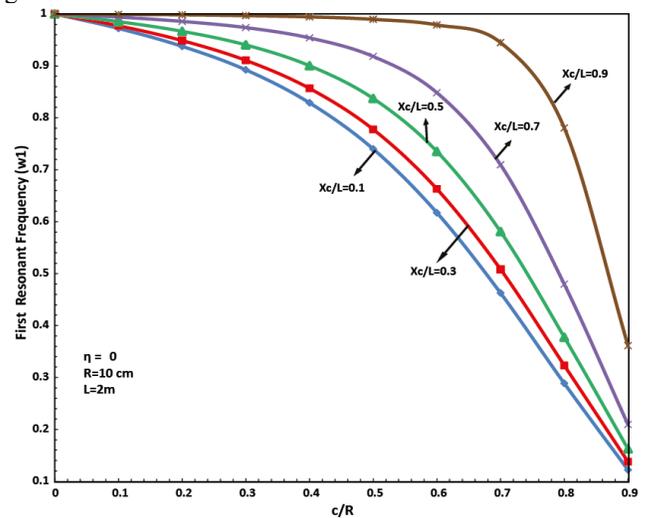

Fig.18. The effect of depth and location of circumferential crack on the first resonant frequency of the shaft.

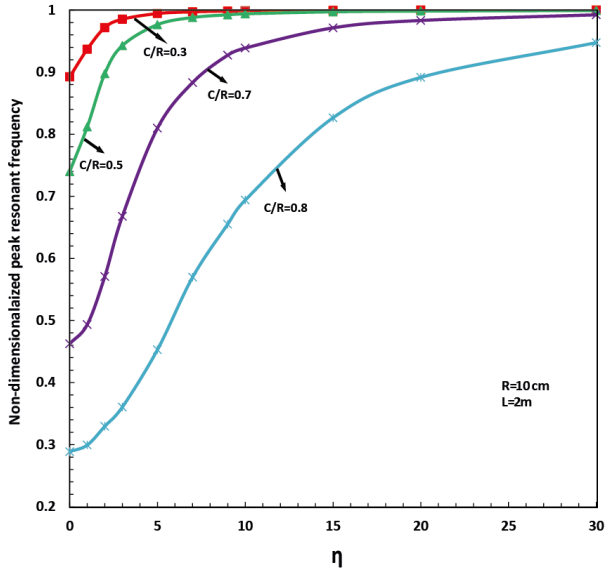

Fig.19. Variation of first non-dimensionalized peak response resonant frequency with energy loss factor

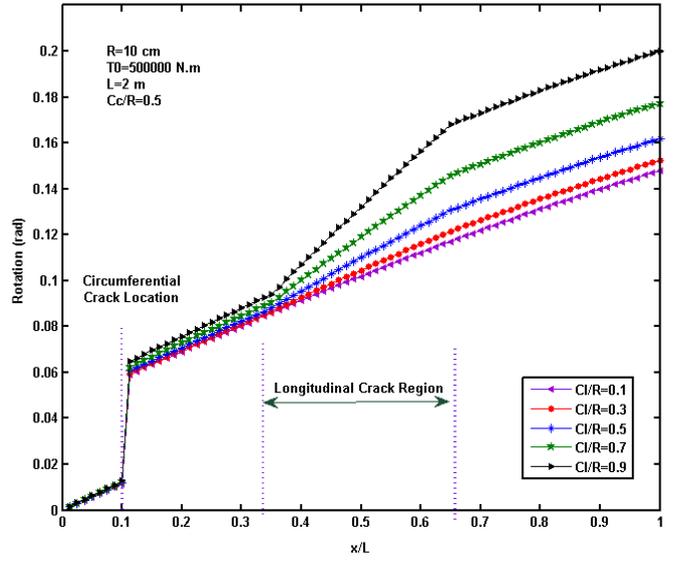

Figure. 20. The indication of rotation along the shaft for a shaft with both circumferential and longitudinal crack with various longitudinal crack depth ratios.

*Case 3) Combination of Longitudinal and Circumferential Crack*

In this part of the research, the dynamic response of a shaft containing both longitudinal and circumferential cracks is investigated. A 2-m long shaft with radius R=0.1 m, with a longitudinal crack at its mid-section was considered for this investigation. The shaft was also considered to contain a circumferential crack either at right or left side of the longitudinal crack. While the length and depth of the longitudinal crack was considered fixed ($\frac{L_L}{L} = 0.3$, $\frac{C_L}{R} = 0.5$) the circumferential crack length and location were systematically changed and their effects on the frequency response were investigated. Fig. 13 shows a typical torsional response of the shaft with both longitudinal and circumferential cracks subjected to harmonic torsion with the magnitude of 500KN.m and frequency of 1000 rad/sec.

Figure 14 shows the effect of circumferential crack length and location on the natural frequency of the shaft with a longitudinal crack. The results show that both crack length and depth have a drastic effect on the shaft first resonance frequency. As the circumferential crack location shift toward the shaft free end, the circumferential crack length has less effect on the shaft resonance frequency and the shaft resonance frequency approaches to that of a shaft with just a longitudinal crack.

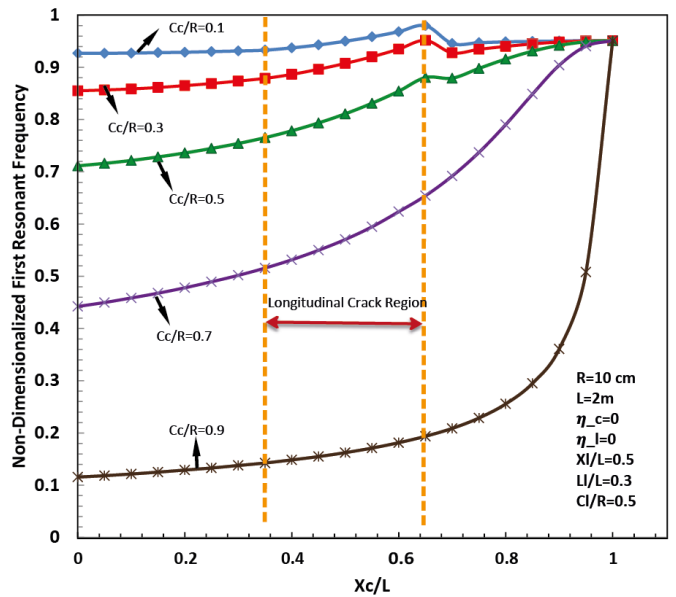

Fig.21. The effect of circumferential crack location and depth on the resonant frequency of a shaft with a longitudinal crack, located in the middle of the shaft.

By changing the circumferential crack location along the shaft, for circumferential crack with the depth greater than the longitudinal crack one Cc/R=0.7, Cc/R=0.9, due to the dominant effect of circumferential crack, figure 21 indicates a smooth and stable behavior in the curves. While, for the circumferential crack with the depth lower than the longitudinal one, a sudden change is observed in the dynamic response of the shaft as circumferential crack passes the longitudinal region approaching the free end. This sudden change does not occur in the clamped side since the shaft dynamic stiffness is exclusively influenced by the circumferential crack located in the clamped end side.

## 3. CONCLUSIONS

Turbo-generator shafts are manufactured through the extrusion process. This results in formation of weak planes in the direction of extrusion. Both longitudinal and circumferential cracks have been observed to form when these shafts are subjected to cyclic torsion. Dynamic response of these shafts could be severely affected by the presence of these cracks. Longitudinal crack section of the shaft is modeled as an equivalent shaft with an adjusted torsional rigidity. The torsional rigidity of the longitudinal cracked shaft was obtained through finite element analysis. It was found the torsional rigidity of shafts with a longitudinal crack can be a function of $\frac{C_L}{R}$. Where the $C_L$ is the crack depth and R is the radius of the shaft. It was observed for $\frac{C_L}{R} < 0.3$, the resonance frequency of the shaft is little affected by the longitudinal crack length. It was found that even when entire shaft have a longitudinal crack, its resonance frequency is little affected. Circumferential cracked shaft is modeled as a shaft having two segments connected by torsional spring. The effect of crack surface interactions was also considered in this investigation. It was found for severe crack surface interactions, the frequency where the peak responses occur, approaches to that of uncracked shafts. The effects of the presence of both longitudinal and circumferential cracks on the shaft resonance frequencies were also investigated. Here again, the circumferential crack location respect to the longitudinal crack dominates the behavior of the shaft.

## ACKNOWLEDGMENTS
We would like to acknowledge Prof. M.Taslim support and encouragement of Mr. Nabian. This research would not have been possible without his support.

## 5. References

1. Ritchie, R.O., McClintock, F.A., Nayeb-Hashemi, H. and Ritter, M.A., 1982. Mode III fatigue crack propagation in low alloy steel. *Metallurgical Transactions A*, *13*(1), pp.101-110.

2. H. Nayeb-Hashemi, F.A. McClintock, and R. O. Ritchie, "Effect of Friction and High Torque on Fatigue Crack Propagation in Mode III," J. of Metallurgical Transaction A, Vol. 13A, Dec, 1982, pp. 2197-2205.

3. H. Nayeb-Hashemi, F. A. McClintock, and R .O. Ritchie, "Influence of Overloads and Block Loading Sequences on Mode Ill Fatigue Crack Propagation in A469 Rotor Steels," Engineering Fracture Mechanics, Vol. 18, No.4, 1983, pp. 763-783.

4. H. Nayeb-Hashemi, S. Suresh, and R. O. Ritchie, "On the Contrast between Mode I and Mode III Fatigue Crack Propagation Under Variable Amplitude Loading Conditions," Material Science and Engineering, 59, 1983, pp. Ll-L5.

5. H. Nayeb-Hashemi, F. A. McClintock, and R. O. Ritchie, "Micro-Mechanical Modeling of Mode III Fatigue Crack Growth in Rotor Steels," Int. J. of Fracture, 23, 1983, pp. 163-185.

6. A. Vaziri and H. Nayeb-Hashemi, "The Effect of Crack Surface Interaction on the Stress Intensity Factor in Mode III Crack Growth in Round Shafts," Engineering Fracture Mechanics, 72, 2005, 617-629

7. A. Vaziri and H. Nayeb-Hashemi, " A Theoretical Investigation on the Vibration Chracteristics and Torsional Dynamic Response of Circumferentially Cracked Turbo-Generator Shafts," Int. J. of Solids and Structures, 43, 2006, pp. 4063-4081.

8. A. C. Ugural and S. K. Fenster, " Advanced Mechanics of Material and Applied Elasticity," Prentice Hall, 2012.

9. T.E.Maselko, "The opening of a nonthrough longitudinal crack in a cylindrical shell taking account of elastoplastic deformation", Journal of Soviet Mathematics, 1993,Vol. 64, pp.953-956.

10. R. D. Adams, P. Cawley, C. J. Pye and B. J. Stone, "A vibration technique for non-destructively assessing the integrity of structures", J. Mechanical Engineering Science, 1978, Vol. 20, pp. 93-100.

11. A. Vaziri and H. Nayeb-Hashemi, "Effects of local energy loss on the dynamic response of a cylindrical bar with a penny shape crack", ASME, 2002, Nov. 17-22 New Orleans, LA.

12. M. Krawczuk and W. Ostachowicz, "Analysis of the effect of cracks on the natural frequencies of a cantilever beam", J. Sound and Vibration, 1991, Vol. 150, pp. 191-201.

13. P. F. Rizos, N. Aspragathos and A. D. Dimarogonas, "Identification of crack location and magnitude in a cantilever beam from the vibration modes", J. Sound and Vibration, 1990, Vol. 138, pp. 381-388.

14. P. G. Kirshmer, "The effect of discontinuities on the natural frequency of beams", Proceedings of ASME, 1994, Vol. 44, pp. 897-904.

15. W. T. Thomson, "Vibration of slender bars with discontinuities in stiffness", J. Applied Mechanics, 1949, Vol. 16, pp. 203-207.

16. H. J. Petroski, "Simple static and dynamic models for the cracked elastic beam", International J. Fracture, 1981, Vol. 17, pp. 71-76.

17. A. Joshi and B. S. Madhusudhan, "A unified approach to free vibration of locally damaged beams having various homogeneous boundary conditions", J. Sound and Vibration, 1991, Vol. 147, pp. 475-488.
18. M-H. H. Shen and C. Pierre, "Free vibrations of beams with a single-edge crack", J. Sound and Vibration, 1994, Vol. 170, pp. 237-259.
19. F. D. Ju and M. E. Mimovich, "Experimental diagnosis of fracture damage in structures by the modal frequency method", ASME, J. Vibration, Acoustics, Stress and Reliability in Design, 1988, Vol. 110, pp. 456-463.
20. P. Cawley and R. D. Adams, "The locations of defects in structures from measurements of natural frequencies", J. Strain Analysis, 1979, Vol. 14, pp. 49-57.
21. C. P. Ratcliffe, "Damage detection using a modified Laplacian operator on mode shape data", J. Sound and Vibration, 1997, Vol. 204, pp. 505-517.
22. A. K. Pandey, M. Biswas and M. M. Samman, "Damage detection from changes in curvature mode shapes", J. Sound and vibration, 1991, Vol. 145, pp. 321-332.
23. K. D. Hjelmstad and S. Shin, "Crack identification in a cantilever beam from modal response", J. Sound and vibration, 1996, Vol. 198, pp. 527-545.
24. A. K. Pandey, M. Biswas, "Damage detection in structures using change in flexibility", J. Sound and Vibration, 1994, Vol. 169, pp. 3-17.
25. S.W. Doebling, L.D. Peterson and K.F. Alvin, "Estimation of reciprocal residual from experimental modal data", American Institute of Aeronautics and Astronautics Journal, 1996, Vol. 34, pp. 1678-1685.
26. M. A. Akgun and F. D. Ju, "Damage diagnosis in frame structures with a dynamic response method", J. Mechanical Structures and Machinery, 1990, Vol. 18, pp. 175-196.
27. O. S. Salawu, "Detection of structural damage through change in natural frequencies: a review", Engineering Structures, Vol. 19, pp. 718-723
28. S. W. Doebling, C. R. Farrar and M. B. Prime, "A summary review of vibration-based damage identification methods", Shock and Vibration Digest, 1998, Vol. 30, No. 2, pp. 91-105.
29. A. D. Dimarogonas, "Vibration of cracked structures: a state of the art review", Engineering Fracture Mechanics, 1996, Vol. 55, pp. 831-857.

30. M Nabian , MT. Ahmadian . Multi-Objective Optimization of Functionally Graded Hollow Cylinders. In ASME 2011 International Mechanical Engineering Congress and Exposition 2011 Jan 1 (pp. 583-590). American Society of Mechanical Engineers.

31. M. Nabian , A. Vaziri, M. Olia , & H. Nayeb-Hashemi, (2013, November). The Effects of Longitudinal and Circumferential Cracks on the Torsional Dynamic Response of Shafts. In ASME 2013 International Mechanical Engineering Congress and Exposition (pp. V04BT04A078-V04BT04A078). American Society of Mechanical Engineers.